\documentclass[12pt,titlepage]{article} \usepackage{epsfig}
\usepackage[round,comma,sort&compress,numbers]{natbib}
\usepackage[bf,small]{caption} \usepackage{amssymb}
\usepackage{xcolor}
\usepackage{multirow}
\usepackage{rotating}

\makeatletter \renewcommand{\@biblabel}[1]{#1.} \makeatother

\newcommand{\bii}{\begin{itemize}}
\newcommand{\eii}{\end{itemize}}
\newcommand{\bie}{\begin{enumerate}}
\newcommand{\eie}{\end{enumerate}}
\newcommand{\beq}{\begin{equation}}
\newcommand{\ee}{\end{equation}}
\newcommand{\eeq}{\end{equation}}
\newcommand{\beqa}{\begin{eqnarray}}
\newcommand{\eea}{\end{eqnarray}}

\newcommand{\lam}{\lambda}

\newcommand{\no}{\noindent}

\newcommand{\al}{\alpha}

\newcommand{\bc}{\begin{center}}
\newcommand{\ec}{\end{center}}

\newcommand{\fref}[1]{Figure\ \ref{fig:#1}}
\newcommand{\tref}[1]{Table\ \ref{tab:#1}}

\newcommand{\eref}[1]{eqn.\ (\ref{eqn:#1})}

\newcommand{\id}{\mathrm{d}}

\newcommand{\dt}[1]
{\mathchoice{\frac{\mathrm{d}#1}
{\mathrm{d}t}}{\mathrm{d}#1/\mathrm{d}t}
{\mathrm{d}#1/\mathrm{d}t}
{\mathrm{d}#1/\mathrm{d}t}}

\newcommand{\mat}{\begin{array}{cc}}
\newcommand{\emat}{\end{array}}

\newcommand{\ra}{\rightarrow}

\newcommand{\LD}{\LD_{50}}

\newcommand{\eps}{\varepsilon}
\newcommand{\ff}{f}
\newcommand{\pp}{\bar d}

\newcommand{\ppa}{\bar d_\alpha}
\newcommand{\pmax}{d_{max}}

\graphicspath{{.}{add_figs/}{figs/}}

\definecolor{light-gray}{gray}{0.6}
\definecolor{dark-gray}{gray}{0.3}

\advance\voffset by -0.8in \advance\hoffset by -0.8in
\renewcommand{\cite}{\citep}

\begin{document} 

\title{Explicit kinetic heterogeneity: mechanistic models for
interpretation of labeling data of heterogeneous cell populations}

\author{{Vitaly V. Ganusov$^{\dag\ddag}$\thanks{Current address:
Theoretical Biology and Biophysics, Los Alamos National Laboratory,
Los Alamos, NM 87545, USA}, Jos\'{e} A.M.  Borghans$^\S$, Rob J. De
Boer$^\ddag$}\\ {\small $^\ddag$Theoretical Biology, Utrecht
University, Padualaan 8, 3584 CH, Utrecht, The Netherlands}\\ {\small
$^\S$University Medical Center Utrecht, Lundlaan 6, 3584 EA Utrecht, The
Netherlands}\\ {\small $^\dag$Institute of Biophysics, Krasnoyarsk,
Russia 660036} }

\maketitle

\begin{abstract} 
  Estimation of division and death rates of lymphocytes in different
  conditions is vital for quantitative understanding of the immune
  system.  Deuterium, in the form of deuterated glucose or heavy
  water, can be used to measure rates of proliferation and death of
  lymphocytes in vivo. Inferring these rates from labeling and
  delabeling curves has been subject to considerable debate with
  different groups suggesting different mathematical models for that
  purpose. We show that the three models that are most commonly used
  are in fact mathematically identical and differ only in their
  interpretation of the estimated parameters.  By extending these
  previous models, we here propose a more mechanistic approach for the
  analysis of data from deuterium labeling experiments.  We construct
  a model of ``kinetic heterogeneity'' in which the total cell
  population consists of many sub-populations with different rates of
  cell turnover.  In this model, for a given distribution of the rates
  of turnover, the predicted fraction of labeled DNA accumulated and
  lost can be calculated. Our model reproduces several previously made
  experimental observations, such as a negative correlation between
  the length of the labeling period and the rate at which labeled DNA
  is lost after label cessation. We demonstrate the reliability of the
  new explicit kinetic heterogeneity model by applying it to
  artificially generated datasets, and illustrate its usefulness by
  fitting experimental data. In contrast to previous models, the
  explicit kinetic heterogeneity model 1) provides a mechanistic way
  of interpreting labeling data; 2) allows for a non-exponential loss
  of labeled cells during delabeling, and 3) can be used to describe
  data with variable labeling length.

\vskip 1cm Classification: Immunology, Mathematical Biology.
  

Short title: Measuring cell turnover

\vskip 0.5cm Abbreviations: D-glucose = $^2H_2$-glucose, heavy water =
$^2H_2O$, KH = kinetic heterogeneity, AM = asymptote model, RSS =
residual sum of squares.

\end{abstract}

\section{Introduction}

There is little consensus about the expected life spans of lymphocyte
populations in health and disease.  Labeling the DNA of dividing cells
with deuterium has proved to be one of the most reliable and feasible
ways to study the population dynamics of lymphocytes in healthy human
volunteers and in patients
\cite{Hellerstein.nm99,Mohri.jem01,Vrisekoop.pnas08}.  Deuterium, in
the form of deuterated glucose or heavy water, is used to measure the
rate at which cells are dividing {\it in vivo}, without the need to
interfere with these cellular kinetics. Deuterium is incorporated into
newly synthesized DNA via the {\it de novo} pathway
\cite{Hellerstein.it99}, and enrichment of deuterium (over hydrogen)
in the DNA of cells is therefore related to cell division.  During
label administration, the fraction of deuterium-labeled nucleotides
increases over time, and after label withdrawal, the fraction
generally declines over time \cite{Mohri.jem01,Vrisekoop.pnas08}.
Labeling DNA with deuterium in humans has a number of clear advantages
over other labeling techniques such as with BrdU, including the
absence of toxicity, the fact that the rate of incorporation of
deuterium into the DNA is independent of the amount of nucleotides
present, and a simpler mathematical interpretation of the data
\cite{Macallan.pnas98,Neese.ab01,Hellerstein.it99}.  Several
mathematical models have been proposed for estimation of cellular
turnover rates from labeling data (reviewed in \cite{Borghans.ir07}),
including a simple precursor-product relationship
\cite{Hellerstein.nm99}, a source model \cite{Mohri.jem01}, a
two-compartment model \cite{Ribeiro.pnas02,Ribeiro.bmb02}, and a
kinetic heterogeneity model \cite{Asquith.ti02}.

In their pioneering study on deuterium labeling, \citet{Mohri.jem01}
found that the estimated rate of cell proliferation was typically
smaller than the rate of cell death. Because the cell population under
investigation was in steady state, the extra death must be compensated
by a source of cells, for example from the thymus. This interpretation
was challenged by the work of \citet{Asquith.ti02}, which pointed out
that estimated proliferation and death rates do not have to be equal
if the population is kinetically heterogeneous (i.e., different cells
in the population divide and die at different rates). %
%
%
Because the labeled population preferentially contains cells that
proliferate (and die) relatively rapidly, the estimated rate of cell
death is in fact expected to be higher than the average proliferation
rate \cite{Asquith.ti02}.

Here we extend these studies and propose a more mechanistic approach
to estimate the rates of lymphocyte proliferation and death from
deuterium labeling experiments.  First, we show that the three most
commonly used mathematical models are in fact kinetically identical
(i.e., will lead to identical estimates of the average rate of cell
turnover), and only differ in their biological interpretation of the
model parameters. Second, we formulate a novel mathematical model
which explicitly takes into account kinetic heterogeneity of
lymphocyte populations, and show how lymphocyte turnover rates can be
calculated using this model.  Several previously made experimental
observations arise naturally from the new model.  For example, we find
that the rate of label loss during delabeling generally exceeds the
rate of label accumulation during the labeling phase. Our model also
explains the dependence of the rate at which labeled DNA is lost after
label withdrawal on the duration of the labeling period
\cite{Asquith.ti02}.
As a proof of principle, we demonstrate that the newly developed model
can fit artificially generated data, and correctly returns their
underlying kinetic parameters. We also illustrate the usefulness of
the new model by fitting it to several experimental datasets.  The
novel explicit kinetic heterogeneity model may offer alternative
interpretations of how infections or treatments affect the turnover of
human lymphocytes {\it in vivo}.

\section{Results}

\subsection{Previous models}

Although different models have been proposed for interpretation of
deuterium labeling data \cite{Mohri.jem01,Asquith.ti02} and are being
debated in the literature, they are in fact mathematically identical.
Following \citet{DeBoer.pbs03}, consider a cell population consisting
of a fraction $\al$ of cells with average turnover rate $d$ (i.e., an
expected life span of $1/d$ days), and a fraction $1-\al$ of cells
that do not turnover at all on the time scale of the experiment.
During the labeling phase, consider the fraction of unlabeled DNA
$U_\al$ in the sub-population with death rate $d$.  Because DNA is
only lost by cell death, $U_\al$ changes according to:

$$\id U_\al/\id t= -d U_\al.$$

During the delabeling phase the fraction of labeled DNA in that same
population ($L_\al$) is described by:
$$\id L_\al/\id t= -d L_\al,$$

\no because labeled DNA can only be lost by cell death.  Since
$U_\al+L_\al=1$, the fraction of labeled DNA in the whole population
$L(t)=\al L_\al(t)$ is described by:

\beq L(t)=\left\{\begin{array}{ll} \alpha\left(1-e^{-dt}\right), &
\mbox{if $t\leq T$},\\ L(T)e^{-d(t-T)}, & \mbox{otherwise},\\
  \end{array} \right. \label{eqn:L-deboer}
\ee

\no where $T$ is the duration of the labeling period. Given that only
a fraction $\al$ of all cells in the population are turning over (or
dying) at rate $d$, the average turnover rate of the whole population
is $\al\times d$ \cite{DeBoer.pbs03}.  Importantly, this approach does
not require us to describe how new cells are formed, i.e., they could
be generated by the thymus and/or by proliferation. Extending this
model by assuming $n$ sub-populations with different rates of cell
proliferation $p_i$ and death $d_i$, and possibly generation of new
cells from a source $s_i$ (\fref{cartoon}), the fraction of labeled
nucleotides in the whole population at time $t$ is given by:

\beq L(t)=\left\{\begin{array}{ll}
\sum_{i=1}^n\alpha_i\left(1-e^{-d_it}\right), & \mbox{if $t\leq T$},\\
\sum_{i=1}^n\alpha_i\left(1-e^{-d_iT}\right)e^{-d_i(t-T)}, &
\mbox{otherwise},\\
\end{array} \right. \label{eqn:L-general}
\ee

\no where $\al_i$ is the fraction of cells in population $i$ with
death rate $d_i$, and $\al=\sum_{i=1}^n\alpha_i\leq 1$ is the
asymptote that would be approached if label would be administered
indefinitely.

The ``source'' model that was previously proposed by
\citet{Mohri.jem01} considered one homogeneous cell population, but
allowed for a source of unlabeled cells during the labeling phase,
i.e., $\dt{U}=s_U-dU$, which also gives rise to an asymptote
$\al=1-\hat s_U/d$, defining the fraction of cells that can maximally
become labeled (here $\hat s_U = s_U/X$ where $X$ is the total number
of cells in the population at equilibrium and $s_U$ is the number of
cells with unlabeled DNA coming from the source per day during the
labeling phase \cite{Mohri.jem01}).  Mathematically, the source model
is therefore identical to \eref{L-deboer}.  Similarly, in the kinetic
heterogeneity model devised by \citet{Asquith.ti02},
$\dt{L}=p(U+L)-dL=p-dL$ for the labeling phase and $\dt{L}=-dL$ for
the delabeling phase. Assuming $p\leq d$ one again obtains
\eref{L-deboer} with $\al=p/d$.  Therefore, all these models are
mathematically identical and only differ in the biological
interpretation of their model parameters (see also
\cite{Asquith.ti09}). We propose to call all these models the
``asymptote model''.  Importantly, in all models the product
$\al\times d$ can be interpreted as the average rate of cell turnover
of the population as a whole \cite{DeBoer.pbs03}, and therefore, all
three models, when fitted to data, will deliver identical estimates of
the average turnover rate, which is the parameter of interest.


\subsection{Kinetic heterogeneity model with continuously distributed
  turnover rates}

\begin{figure}[e]
\bc
\includegraphics[width=0.9\textwidth]%
{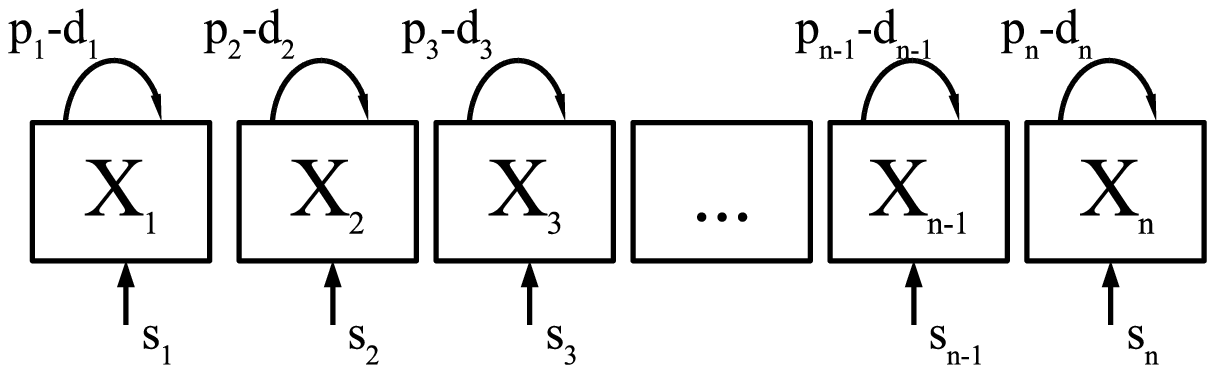} \ec
\caption{A cartoon of the model with explicit kinetic heterogeneity.
  In the model, the population of cells consists of $n$
  sub-populations with different rates of turnover. In the $i^{th}$
  sub-population, there is a source of new cells that enter the cell
  population at rate $s_i$ cells per day, cells divide at rate $p_i$
  per day, and die at rate $d_i$ per day. To maintain the size of
  all sub-populations constant, $d_i - p_i = s_i/X_i$ for every
  sub-population $i$, where $X_i$ is the number of cells in the
  $i^{th}$ sub-population. In this model we assume that the source
  produces only labeled cells during the labeling phase, and 
  delabeled cells during the unlabeling phase \cite{DeBoer.pbs03}.  }
\label{fig:cartoon}
\end{figure}

Because of its simplicity, the model given in \eref{L-deboer} has two
limitations. First, the asymptote level is a phenomenological
parameter that depends on the length of the labeling period
\cite{Asquith.ti02}.  As a consequence, datasets with different
labeling periods will likely give rise to different estimated
asymptotes and different estimated average rates of cell turnover.
Therefore strictly speaking, this model cannot be used to explain
multiple datasets coming from the same experimental setup varying only
in the length of the labeling period; the differences in the rate at
which labeled DNA is lost would force either the asymptote or the
estimated average turnover rate to be different for the different
labeling periods (Den Braber et al. in preparation). Second, the model
assumes that the increase in labeled DNA during the uplabeling phase,
and the loss of labeled DNA during the delabeling phase can be
described by single exponentials. This may be incorrect if cell
populations with different turnover rates are labeled and subsequently
lost.

Under very general assumptions, we have formulated an
alternative model that does not make these {\it a priori} assumptions.
In our new model, a cell population consists of $n$ sub-populations
each with different kinetic properties (see \fref{cartoon} and
\eref{L-general}). If the number of sub-populations is
large ($n\ra\infty$), the sum in \eref{L-general} can be replaced by
an integral. The fraction of labeled nucleotides in the population
then becomes (see Supplemental Information for derivation)

\beq L(t) = \int_0^\infty L(t,d)\id d \label{eqn:L-int}\ee

\no where $L(t,d)$ is given by \eref{L-deboer} where $\al=\ff(d)$ is
the frequency distribution of turnover rates, and $\ff(d)\id d$ is the
probability that a randomly chosen cell in the population belongs to a
sub-population with a turnover rate in the range $(d,d+\id d)$.  If
the turnover rates in the population, $\ff(d)$, follow a gamma
distribution, the change in the fraction of labeled DNA with time is
given by:

\beq L(t)=\left\{\begin{array}{ll} 1-\left(1+\frac{\bar{d} t}{k}\right)^{-k}, &
\mbox{if $t\leq T$},\\ \left(1+\frac{\pp (t-T)}{k}\right)^{-k}-\left(1+\frac{\pp
t}{k}\right)^{-k}, & \mbox{otherwise},\\
\end{array} \right. \label{eqn:L-gamma}
\ee

\no where $\pp$ is the average rate of cell turnover in the
population, $k$ is the shape parameter of the gamma distribution, and
$T $ is the duration of the labeling period.  For $k=1$, the gamma
distribution becomes an exponential distribution, and the rate at
which the fraction of labeled DNA changes is simply:

\beq L(t)=\left\{\begin{array}{ll} {\pp t\over 1+\pp t}, & \mbox{if
$t\leq T$},\\ {\pp T\over (1+\pp t)(1+\pp (t-T))}, &
\mbox{otherwise}.\\
\end{array} \right. \label{eqn:L-exp}
\ee

This is an interesting model in which a single parameter $\pp$
predicts both the rate of uplabeling and downlabeling, and in which
there is no asymptote below 100\% for the level of labeled DNA, i.e.,
under continuous label administration all cells in the population will
become labeled (\fref{chase}). Moreover, this model predicts that the
initial rate $d^*$ at which labeled DNA is lost after label cessation
depends on the duration of the labeling period, $d^*\approx
\pp(1+(1+\pp T)^{-1})$ (see Supplementary Information for derivation).
According to this model, short labeling experiments ($\pp T\ll 1$,
$d^*\approx 2\pp$) will lead to 2-fold faster initial rates of decline
in the fraction of labeled nucleotides than longer labeling
experiments ($\pp T\gg 1$, $d^*\approx\pp$).

\begin{figure}
\includegraphics[width=0.99\textwidth]{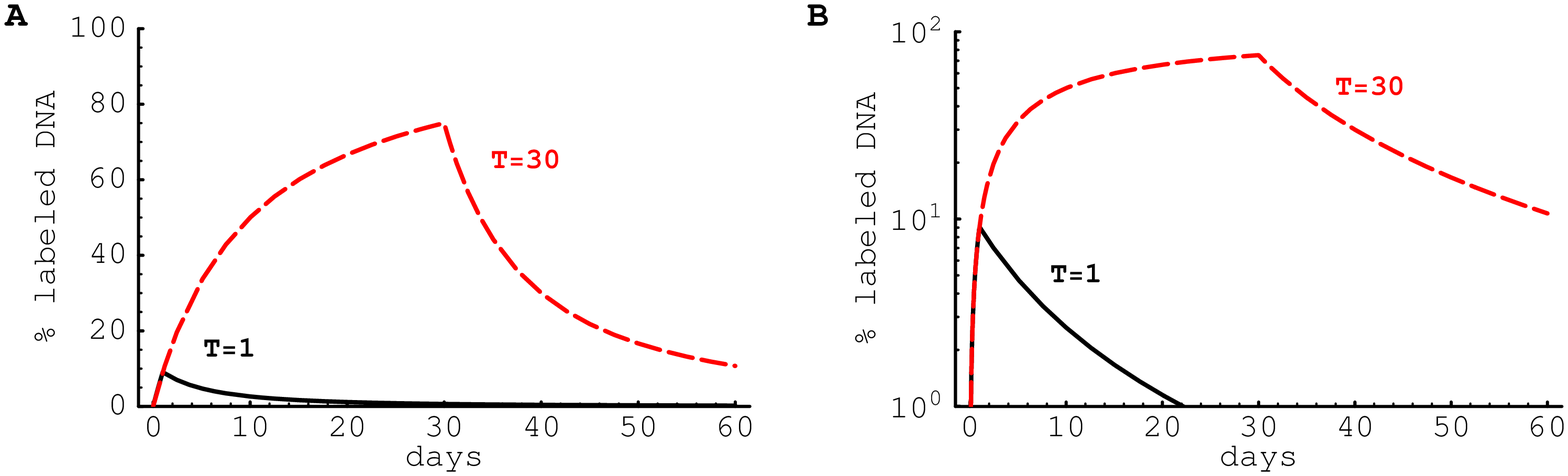}
\caption{Model predictions for exponentially distributed turnover
  rates.  We have plotted the changes in the fraction of labeled DNA
  according to the explicit kinetic heterogeneity model with
  exponentially distributed turnover rates (\eref{L-exp}, mean $\pp
  =0.1/$day).  Predicted changes are shown for a short labeling period
  ($T=1\ $day, solid line) and a long labeling period ($T=30\ $days,
  dashed line) on a linear (panel A) and a logarithmic (panel B)
  scale.  The initial uplabeling rate is independent of the length of
  the labeling period and is given by $\pp$.  The initial rate of
  delabeling, in contrast, depends on the length of the labeling
  period and is approximately twice as fast in the case of short-term
  labeling as compared to long-term labeling.}
\label{fig:chase}
\end{figure}

\begin{figure}
\includegraphics[width=0.99\textwidth]{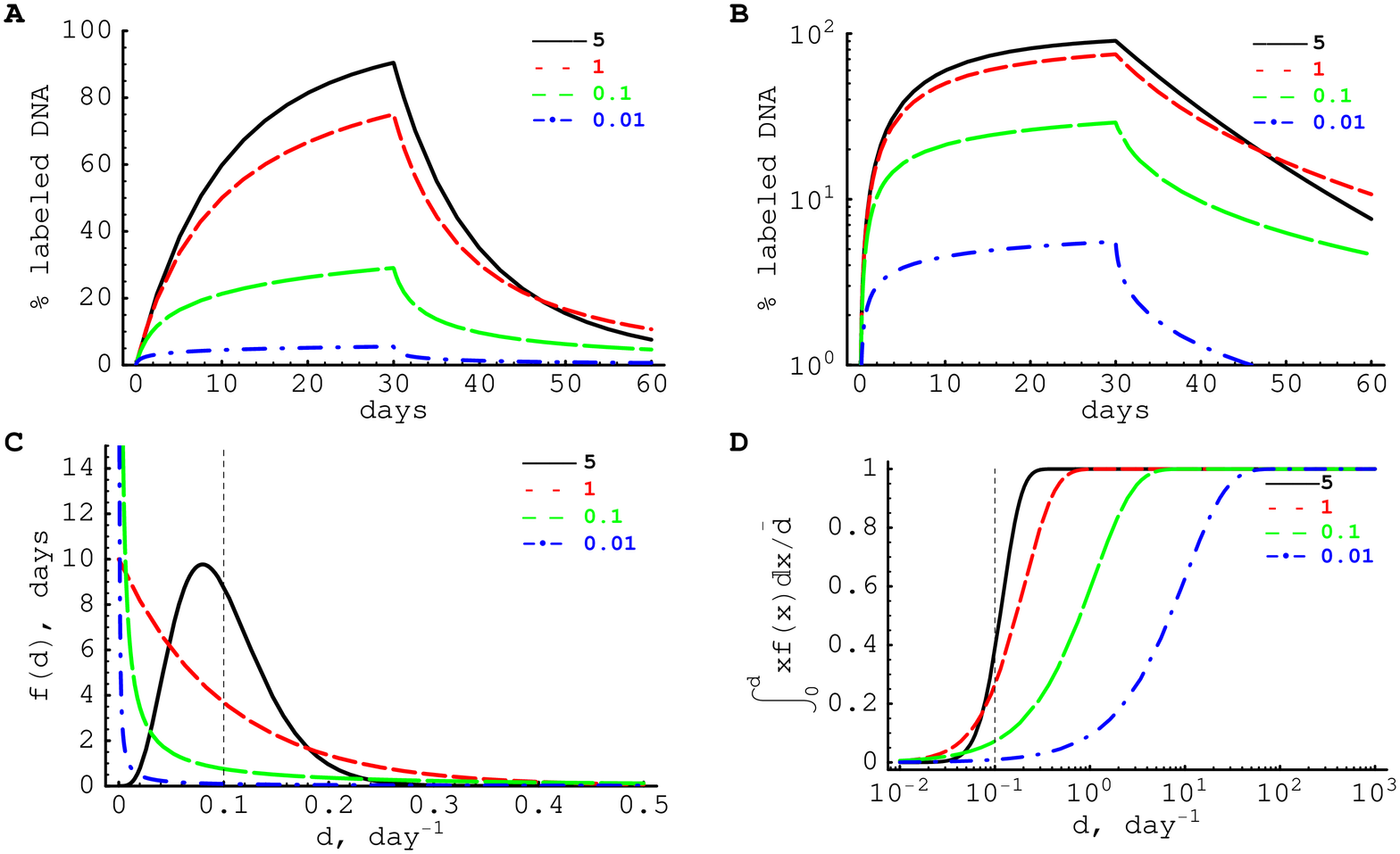}
\caption{Model predictions for gamma-distributed turnover rates. 
We have plotted the changes in the fraction of labeled DNA according to the kinetic
heterogeneity model with gamma-distributed turnover rates
(\eref{L-gamma}) with average turnover rate $\pp =0.1$/day on a linear
(panel A) or logarithmic (panel B) scale.  Predicted changes are shown
for different values of the shape parameter $k$. Larger values of $k$
correspond to a more symmetric distribution (Panel C). For
low values of the shape parameter $k$, the loss of labeled DNA after
label cessation is biphasic, which is most clearly visible on a
logarithmic scale for $k<1$ (panel B). This characteristic of the
kinetic heterogeneity model differs from the Asymptote models which
have a constant {\it per capita} rate at which labeled DNA is
lost. Note that for shape parameters $k<1$, the distribution of
turnover rates $\ff(d)$ becomes extremely skewed with most cells
undergoing hardly any division and relatively few cells undergoing
extremely many rounds of division (panel C). Panel D gives the
cumulative contribution of sub-populations with a particular turnover
rate $d$ to the average rate of turnover of the population $\pp$. The
vertical line shows the value of the average proliferation rate
$\pp$. For high values of the shape parameter ($k=5$), the cell
sub-populations with turnover rates that are somewhat lower or higher
than $\pp$ give the main contribution to the average turnover rate.
In contrast, for low values of $k$ ($k=0.01$), the major contribution
to the average turnover rate comes from sub-populations with extremely
rapid turnover rates ($d\gg\pp$); about 50\% of the average turnover
is due to a few sub-populations with turnover rates that exceed 10 per
day, which is biologically unrealistic.
}
\label{fig:gamma}
\end{figure}

Solutions (\ref{eqn:L-gamma}) and (\ref{eqn:L-exp}) predict that the
initial rate of increase in the fraction of labeled DNA is the average
rate of cell turnover $\pp$ (see also Supplementary Information).
However, the increase in the fraction of labeled DNA does not appear
to be exponential, as was implicitly assumed in the asymptote models
discussed above.  Similarly, during the delabeling period, the model
predicts a non-exponential decline in the fraction of labeled DNA
(\fref{chase} and \ref{fig:gamma}). In general, the initial rate of
label loss during delabeling $d^*$ is given by:

\beq d^* \approx \pp + {\mbox{var}(d)\over \pp},\label{eqn:d*}\ee

\no where $\mbox{var}(d)$ is the variance of the distribution of
turnover rates in the population. In case when turnover rates follow a
gamma distribution, the initial rate of loss of the label after short
labeling periods depends on the shape parameter $k$ of the
distribution, $d^* \approx \pp(1+1/k)$, while it does not after long
labeling periods ($d^*\approx \pp$). The rate of loss of labeled DNA
slows down as less DNA remains labeled, which is most clearly seen
when proliferation rates are distributed according to a very skewed
gamma distribution ($k<1$, \fref{gamma}).  This is a natural property
of the explicit kinetic heterogeneity model as loss of labeled DNA is
reflecting the distribution of the turnover rates of the different
sub-populations, with labeled DNA from the most rapidly turning over
sub-populations being lost first (early fast decline) and labeled DNA
from the other, more slowly turning over, populations being lost later
(late slow decline).

To study the effect of the shape of the turnover rate distribution on
the predicted labeling curve, we plotted the changes in the fraction
of labeled DNA as predicted by the model (\fref{gamma}A\&B) with
different gamma-distributed turnover rates (\fref{gamma}C).  When the
gamma distribution is highly skewed (i.e., $k<1$), the majority of
cell sub-populations have very low rates of cell turnover, and the
average rate of cell turnover is dominated by a few sub-populations
that turn over unrealistically fast. This is best illustrated by
calculating the cumulative contribution of a sub-population with a
particular rate of turnover to the average turnover rate of the
population $\pp$ (\fref{gamma}D):

\beq \beta(d)={1\over \pp} \int_0^d x\ff(x)\id x. \ee

For large values of the shape parameter $k$ (e.g., $k=5$), the
sub-populations with turnover rates that are close to the average
turnover rate $\pp$, are the main contributors to the average rate of
cell turnover (\fref{gamma}D).  When the gamma distribution is
extremely skewed ($k=0.01$), the rate of turnover of the
sub-populations that contribute significantly to the average turnover
rate is as high as $10^1-10^2$ per day, which is biologically
unrealistic.  Therefore, the gamma distribution should be rejected
whenever one estimates a high average turnover rate $\pp$ with a
highly skewed gamma distribution (i.e., a low value of the shape
parameter $k$). As a rule of thumb, $k$ should be larger than 0.1
(\fref{gamma}D); otherwise a relatively large fraction of
sub-populations has unrealistically high turnover rates.

It is possible, however, that not all cells in the population are
turning over.  The models above can easily be extended to incorporate
this possibility by allowing for the same asymptote as in
\eref{L-deboer}.  An example would be a labeling experiment in which
slowly turning over naive T lymphocytes and more rapidly turning over
memory lymphocytes are not separated \cite{Mohri.jem01}. If only a
fraction $\al$ of cells have turnover rates that are distributed
exponentially, and the other cells undergo negligible turnover on the
time scale of the experiment, the change of the fraction of labeled
nucleotides with time is given by:

\beq L(t)=\left\{\begin{array}{ll} {\alpha \ppa t\over 1+\ppa t}, &
    \mbox{if $t\leq T$},\\ {\alpha \ppa T\over (1+\ppa t)(1+\ppa
      (t-T))},
    & \mbox{otherwise},\\
\end{array} \right. \label{eqn:L-expa}
\ee

\no where $\ppa$ is the average of the exponentially distributed
turnover rates, and the average rate of cell turnover in the whole
population is $\pp=\al\ppa$.

It should be noted that the results of this section are applicable
both to proliferating and non-proliferating lymphocytes, given the
general structure of the cell population in the model (see
\fref{cartoon}). As a downside of this, the model does not allow to
estimate which fraction of labeling of lymphocytes is due to
proliferation of precursors (e.g., thymocytes for naive T cells) or
due to peripheral proliferation of the lymphocyte population itself.
Additional experiments, such as thymectomy in case of studies of naive
T-cell turnover, may allow to estimate the separate contribution of
peripheral T-cell proliferation \cite[Den Braber et al.
submitted]{Parretta.ji08}.


\subsection{Fitting artificial data to validate the model}

Having analytical expressions for several kinetic heterogeneity
models, we analyzed how well these models can recover the (known)
average turnover parameter from simulated (artificial) datasets. Three
models were used to generate artificial datasets: 1) the kinetic
heterogeneity model with gamma-distributed rates of turnover
(\eref{L-gamma}), referred to as the ``Gamma model'', 2) the kinetic
heterogeneity model in which a fraction $\al\leq 1$ of cells have
exponentially-distributed rates of turnover (\eref{L-expa}), referred
to as the ``Exponential model'', and 3) a ``Two population model''
(\eref{L-general} with $n=2$, turnover rates $d_1$ and $d_2$, average
turnover rate $\pp = \al d_1 + (1-\al)d_2$, and $d_1>\pp$). These
datasets were subsequently fitted by the same three models as well as
by the conventional Asymptote model (\eref{L-deboer}).

Not surprisingly, the models delivered correct estimates for the
average turnover parameter if a dataset was fitted with the model that
was used to generate the data (\tref{simdata} and \fref{simdata_p};
with the results obtained by fitting the Two population model not
shown).  All models described the data sets generated by the other
models reasonably well (\fref{simdata}), although some features in the
data could not be reproduced. For example, the Asymptote model failed
to describe the decreasing rate at which labeled DNA is lost over
time, which is observed in the data generated by the Gamma and the
Exponential models (see last data points in \fref{simdata}). Some
model fits delivered incorrect estimates for the average turnover rate
if the data were generated using another model. For example, the Gamma
model overestimated the average turnover rate when the data were
generated using the Exponential model (up to 2-fold), and
underestimated $\pp$ for data generated using the Two populations
model (over 2-fold). This is most likely due to the strong constraint
of the model that both uplabeling and delabeling curves have to be
described with one mechanism, i.e., gamma-distributed turnover rates.
On the other hand, the Asymptote model always underestimated the true
average turnover rate (up to 2-fold for data generated by the
Two-populations model; \tref{simdata}). It did perform somewhat better
than the Gamma model as judged by the mean square distances, because
the rate of uplabeling and downlabeling are relatively independent in
the Asymptote model.

Given that natural lymphocyte populations are likely to contain
resting sub-populations, some extent of saturation in the fraction of
deuterium-labeled nucleotides is expected in almost any experimental
dataset. In our artificial data, such an asymptote was imposed when
using the Exponential model by letting only 50\% of all cells to turn
over (\fref{simdata} and \tref{simdata}). It is therefore not
surprising that the Gamma model, which does not have an explicit
asymptote in the uplabeling phase (see \eref{L-gamma}), did not
correctly estimate the average turnover rate for the data generated by
the Exponential model (\fref{simdata_p}).  Extending the Gamma model
to allow for an explicit asymptote during the labeling phase ($\al$)
indeed improved the estimate of the average turnover rate $\pp=\al\ppa
= 0.13\ {\rm day}^{-1}$ (with 95\% CIs = $0.095-0.22\ {\rm day}^{-1}$
which includes the true average $\pp=0.1\ day^{-1}$), even though the
estimated fraction of turning over cells $\al$ was not significantly
different from 1 (i.e., an F-test would not reject a model with
$\al=1$; results not shown). This exercise illustrates that when
fitting experimental data one should check whether allowing for an
explicit asymptote in the uplabeling phase leads to different
estimates of the average turnover rate.  Interestingly, all models
underestimated the average rate of cell turnover when the data were
generated using the Two populations model.  This is because the models
did not reproduce the relatively rapid accumulation of the labeled DNA
in the first days (\fref{simdata}).  Fitting the Two populations model
to these data led to better estimates of the average turnover rate
($\pp=0.084$ per day with 95\% CIs = $(0.062-0.35)$ for 7 days of
labeling, and $\pp=0.26$ per day with 95\% CIs = $(0.061-0.44)$ for 15
days of labeling, where the constant $\pp=0.1\ day^{-1}$ is contained
within both ranges, results not shown).

Although stable isotope labeling seems to be the best tool at hand to
estimate rates of lymphocyte turnover, a recent review
\cite{Borghans.ir07} pointed out that estimated lymphocyte turnover
rates differ consistently, depending on the labeling method used
(heavy water or deuterated glucose), and the length of the labeling
period.  {\it A priori}, according to the Asymptote model that is
generally used, the estimated average turnover rate should not depend
on the length of the labeling period. Using our explicit kinetic
heterogeneity models, we analyzed the influence of the length of the
labeling period on the estimated average turnover rate. For all
models, we found that the duration of labeling had little influence on
the estimated average turnover rate (for the chosen labeling periods
of 7 and 15 days, \tref{simdata}).  This suggests that longer labeling
periods will not necessarily result in lower estimates of the average
cell turnover rate than shorter labeling periods.

An overall conclusion of this analysis is that without a good
understanding of the underlying model of cell proliferation (i.e., the
distribution of turnover rates in the population), one may obtain
incorrect estimates of cellular turnover rates, even if the quality of
the fit to the data is acceptably good.  Therefore, when analyzing
experimental data, one should aim at using several alternative models
for fitting, and investigate whether estimates of kinetically
important parameters, such as the average rate of cell turnover, are
independent of the model used.

\begin{figure}
  \bc
7 days labeling \hspace{5cm} 15 days labeling\\
  \includegraphics[width=0.99\textwidth]{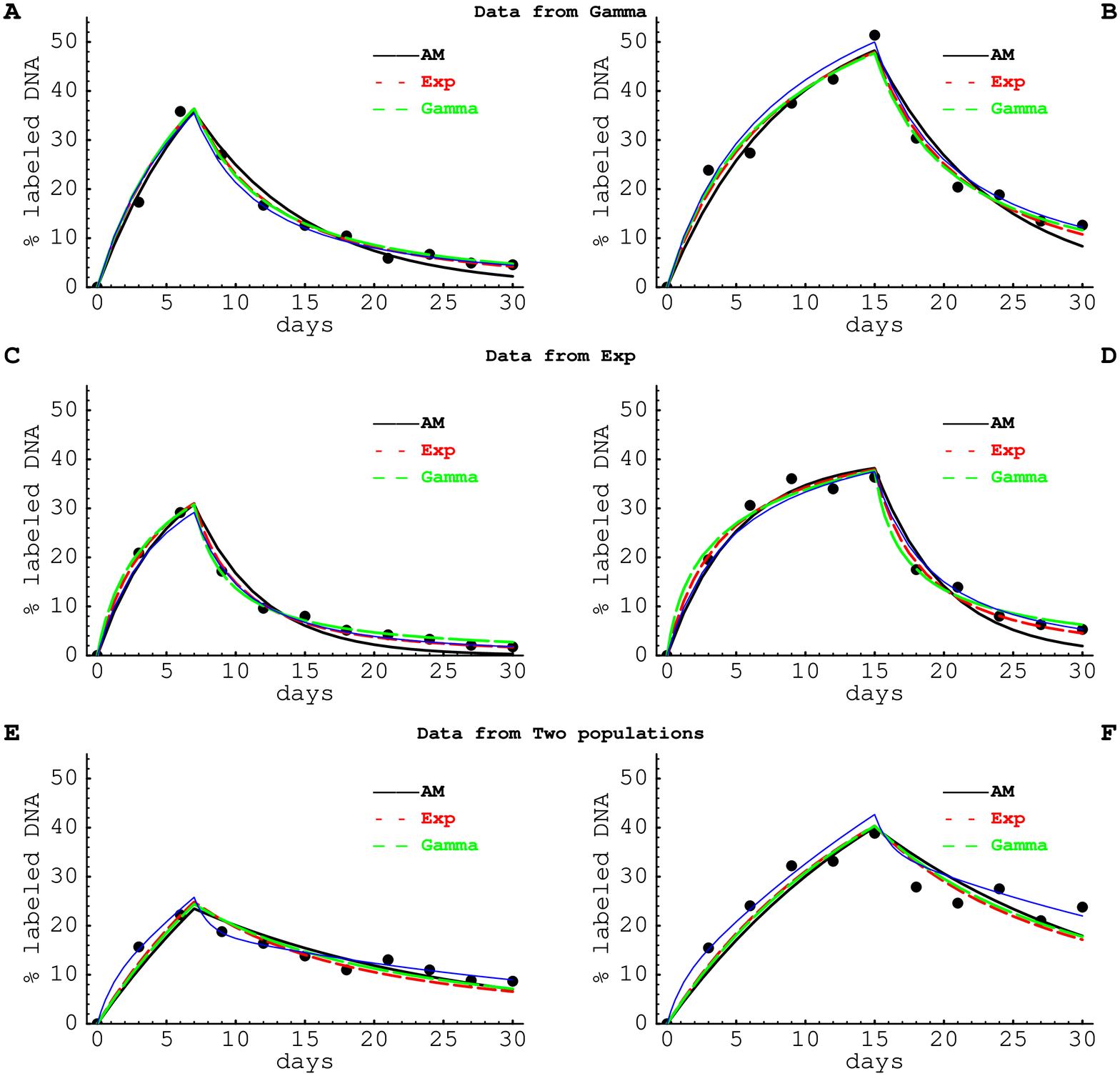}
\end{center}
\caption{Fitting artificial data (black dots) with the Asymptote model
  (\eref{L-deboer}, solid black lines), the Exponential model, in
  which a fraction $\alpha$ of the cells have exponentially
  distributed turnover rates (\eref{L-expa}, small red dashed lines),
  and the Gamma model with gamma distributed turnover rates
  (\eref{L-gamma}, large green dashed lines).  Data were generated
  using the Gamma model (panel A\&B), the Exponential model (panel
  C\&D) and the Two-populations model (panel E\&F, \eref{L-general}),
  respectively.  Thin blue lines show the exact curves of the models
  that were used to generate the data. The different models were
  fitted to 11 datapoints taken from these predicted curves after
  having added noise to these data points. Noise was added by a
  relative change of the predicted value with a normally distributed
  error (with standard deviation of the distribution $\sigma=0.1$).
  The models were fitted to data from artificial labeling experiments
  in which the label was administered for 7 (left panels) or 15 (right
  panels) days.  Parameter estimates providing the best fit are shown
  in \tref{simdata}, and the corresponding estimates of the average
  rates of cell turnover $\pp$ are shown in \fref{simdata_p}.
  Parameters used to generate the data are also given in
  \tref{simdata}.  }
\label{fig:simdata}
\end{figure}

\begin{figure}
  \bc
\includegraphics[width=0.99\textwidth]{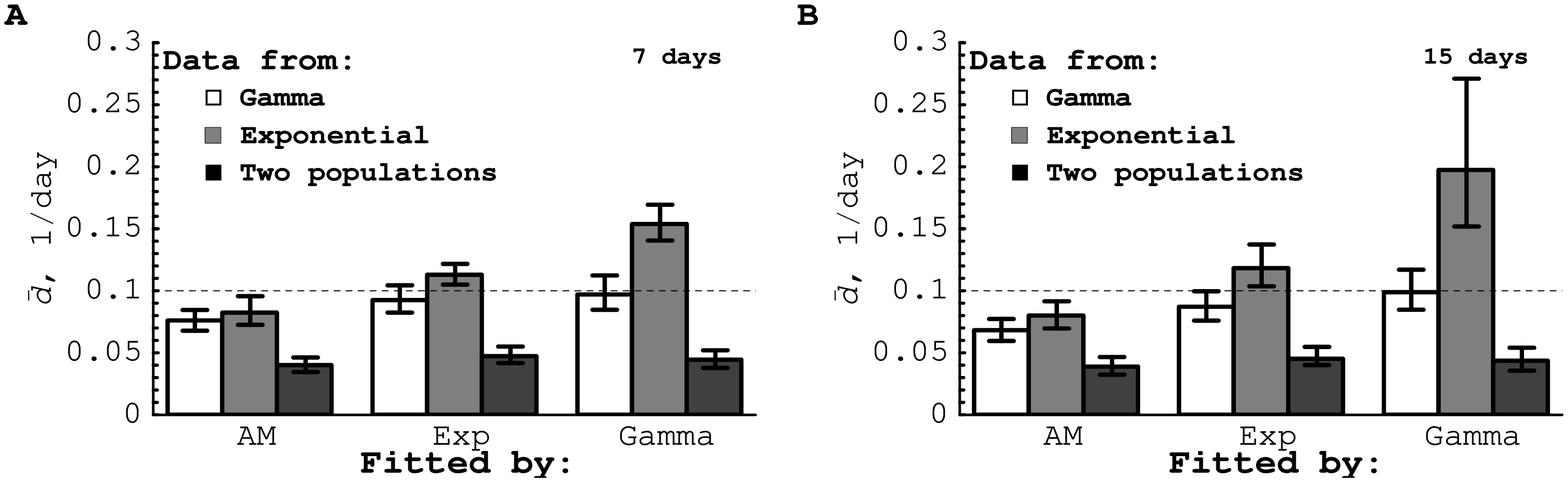}
\end{center}
\caption{By fitting the artificial data described in the text, we
  estimated the average turnover rate using three models: the
  Asymptote model (\eref{L-deboer}), the Exponential model (in which a
  fraction $\alpha$ of cells have exponentially distributed turnover
  rates, \eref{L-expa}), and the Gamma model (with gamma-distributed
  turnover rates, \eref{L-gamma}).  Estimated mean values and 95\%
  confidence intervals obtained by bootstrapping the residuals with
  1000 simulations are shown. Data were generated using the Gamma
  model ($\Box$), the Exponential model
  (\textcolor{light-gray}{$\blacksquare$}) and the Two-populations
  model (\textcolor{dark-gray}{$\blacksquare$}). Labeling periods were
  7 (panel A) and 15 (panel B) days. Horizontal dashed lines denote
  the actual average rate of lymphocyte turnover in all data,
  $\pp=0.1$/day. Note that in this example, the Asymptote model always
  underestimated the average rate of cell turnover, and that there is
  a systematic 2-fold underestimation of the average turnover by all
  models when the data from the Two-populations model 
  were fitted. This is because
  all three models fail to describe the relatively rapid
  accumulation of the label at early time points (see
  \fref{simdata}E--F).
}
\label{fig:simdata_p}
\end{figure}

\subsection{Fitting experimental data}

We next sought to determine how well the new kinetic heterogeneity
models fit experimental data.  Using deuterated glucose,
\citet{Mohri.jem01} obtained labeling data of T lymphocytes from
uninfected healthy human volunteers and from chronically HIV-infected
patients. Previously, these data were fitted using an extended
4-parameter source model, to estimate the rates of cell
division and death of T lymphocytes in healthy humans, and to obtain insights
into how these rates change upon HIV-infection \cite{Mohri.jem01}.
Lymphocytes were sorted into CD4$^+$ and CD8$^+$ T cells, without
distinguishing between their naive and memory subpopulations. Since
naive T cells have a much slower rate of turnover than memory T cells
\cite{Vrisekoop.pnas08}, it is natural to assume an asymptote in the
fraction of labeled nucleotides of unsorted CD4$^+$ and CD8$^+$ T
cells.

We have refitted the labeling data from the four healthy human
volunteers studied by \citet{Mohri.jem01}, again using the three
models for cell proliferation: the Asymptote model (\eref{L-deboer}),
the Exponential model (with a fraction $\alpha$ of cells with
exponentially distributed turnover rates, \eref{L-expa}) and the Gamma
model (with gamma distributed turnover rates, \eref{L-gamma}).  The
data were fitted simultaneously for all four healthy volunteers while
searching for the minimal number of parameters that describe the data
with reasonable quality (using a partial F-test for nested models
\cite{Bates.b88}).  Overall, the models described the data reasonably
well (\fref{mohri.controls} and \tref{mohri.controls.cd4} and
\ref{tab:mohri.controls.cd8}). For CD4$^+$ T cells, the average
turnover rate and the delay at which labeled cells appeared in the
blood did not differ significantly between patients
(\tref{mohri.controls.cd4}).  For all patients, the average rate of
turnover was about 0.46\% per day with a corresponding estimated
half-life of $\ln 2/\pp \approx 151$ days.  There was an average delay
of one day before labeled cells appeared in the blood.  The average
turnover rate of CD4$^+$ T cells from control c1 was always higher
than that of the other individuals, irrespective of the model used
(\fref{mohri.controls.p}A), which may be a sign of an immune response
to an infection in c1 (see also below).  Both the Asymptote model
($\al\approx 15\%$) and the Exponential model ($\al\approx 25\%$)
predicted an asymptote in labeling that is smaller than the fraction
of memory phenotype CD$4^+$ T cells in humans of that age
\cite{Vrisekoop.pnas08}.  The Gamma model could describe these data
even better than the other two models with no need for an asymptote.
The observation that the estimate of the asymptote can differ
dramatically between different models reconfirms our statement that
this parameter is of little use for data interpretation
\cite{DeBoer.pbs03}.

For CD8$^+$ T cells, the parameters differed significantly between
different healthy volunteers, with the exception of the asymptote
level $\alpha$ in the Exponential model which could be fixed between
individuals. The estimates of the average turnover rates of CD8$^+$ T
cells in healthy volunteers c2--c4 did not strongly depend on the
model that was used to fit the data. However, the estimated turnover
rate of CD8$^+$ T cells in individual c1, which was much higher than
the estimated turnover rate in the other healthy volunteers, depended
strongly on the model used and was estimated to be the highest when
using the Gamma model. The latter model fitted the labeling data from
all four individuals very well and reproduced the non-exponential
change in the fraction of labeled DNA in the downlabeling phase
(\fref{mohri.controls}F).  In all four healthy individuals, CD8$^+$ T
cells turned over at a slower rate than CD4$^+$ T cells; the average
turnover rate of CD8$^+$ T cells was $\pp = 0.29\% $ per day with a
corresponding half-life of $\ln2/\pp\approx 239\ {\rm days}$.  The
fits of the Asymptote model and the Exponential model predicted an
asymptote in labeling of $\alpha\approx 0.15$
(\tref{mohri.controls.cd8}).  Even though the Gamma model lacks an
explicit asymptote lower than 1, it fitted these data with equally
good quality as the models with explicit asymptotes. Allowing for an
explicit asymptote in the Gamma model did not improve the quality of
the fit (CD4$^+$ T cells: $p>0.38$, CD8$^+$ T cells: $p>0.99$,
F-test), and the estimated average lymphocyte turnover rates were not
affected by the addition of an explicit asymptote (results not shown).


It is important to investigate whether the good description of the
data of the model with gamma distributed turnover rates is achieved
with biologically reasonable parameter values. In all data we
estimated the shape parameter of the gamma distribution to be small,
i.e. $k<1$, but $k$ was estimated to be larger than 0.1 in seven of
the eight fits. Low values of the shape parameter $k$ imply that in
the population most cells turn over at very slow rates while a few
populations turn over very rapidly.  To investigate whether such a
distribution is biologically reasonable, we calculated the fraction of
cells in the population with a turnover rate higher than $\pmax=1$ per
day which is the maximal rate of CD8$^+$ T-cell proliferation in
rhesus macaques \cite{Davenport.jv04}. This fraction is given by
$\int_{\pmax}^\infty \ff(d)\id d$ for the estimated parameters of the
distribution (see \tref{mohri.controls.cd4} and
\ref{tab:mohri.controls.cd8}). For most fits, the fraction of cells
with turnover rates higher than 1 per day is $\leq 10^{-11}$, and
given the estimated total number of lymphocytes in humans of $\sim
10^{12}$ \cite{Ganusov.ti07}, that would yield only a few cells with
unrealistically high rates of turnover.  However, for the CD8$^+$ T
cells of healthy volunteer c1 we found that $\sim
3\times10^{-5}\times10^{12}=3\times10^7$ cells turn over at rates
higher than 1 per day, which is unrealistically high.

To investigate this further we reanalyzed the CD8$^+$
T-cell labeling data of individual c1 using several extended models.
In the first model, a fraction $\al$ of cells in the population have
gamma-distributed turnover rates while the other fraction ($1-\al$) of
cells turn over at the highest possible rate $\pmax=1\ day^{-1}$. This
situation may correspond to a scenario where a small fraction of
CD8$^+$ T cells is responding to an infection. However, this model
failed to describe the data with biologically reasonable parameter
values ($\al\approx 1$ and $k=0.03$).

In the second extended model, the gamma distribution of turnover rates
was truncated at a maximal value $\pmax = 1$ (see Supplementary
Information for analytical results). The fit of this model to the
labeling data for individual c1 was of similar quality as the fit in
which the gamma distribution was not truncated, and it delivered
similar estimates for the average turnover rate and the shape
parameter ($\pp = 0.66\%$ per day and $k=0.030$, results not shown).
We estimate that in healthy volunteer c1 about 0.1\% of all CD8$^+$ T
cells are turning over rapidly at rates between $0.5-1.0$ per day,
which is reasonable.  For example, in mice responding to lymphocytic
choriomeningitis virus (LCMV) infection, at the peak of the immune
response more than 50\% of all CD8$^+$ T cells in the spleen are
specific for the virus \cite{Murali-Krishna.i98,Homann.nm01}.

Finally, in the third model, we assumed that the CD8$^+$ T-cell
population in patient c1 consists of naive, memory and effector T-cell
subpopulations with 3 different rates of turnover (see
\eref{L-general} with $n=3$). Assuming that the rate of turnover of
naive T cells is 0 and that letting for effector cells $\pmax=1$ per
day, we could obtain excellent fits of the labeling data with an
estimated average turnover rate $\pp=1.4\%$ per day ($95\%\mbox{
  CIs}=(1.2-1.6)\%$ per day) which is much higher than estimates
obtained by other models (\fref{mohri.controls.p}). Using model
selection methods such as the Akaike Information Criterion, we found
equal support for the latter model and the model in which the turnover
rates follow a truncated gamma distribution \cite[results not
shown]{Burnham.b02}. We can conclude, therefore, that the average
turnover rate of CD8$^+$ T cells in patient c1 is at least 0.62\% per
day (Gamma model) and could be as high as 1.4\% per day (Three
population model).  In summary, it seems that the average turnover
rate of both CD4$^+$ and CD8$^+$ T cells was increased in individual
c1 as compared to other individuals, and this could be explained by a
normal immune response in this otherwise healthy volunteer.


\begin{figure}
\bc 
 \includegraphics[width=0.99\textwidth]{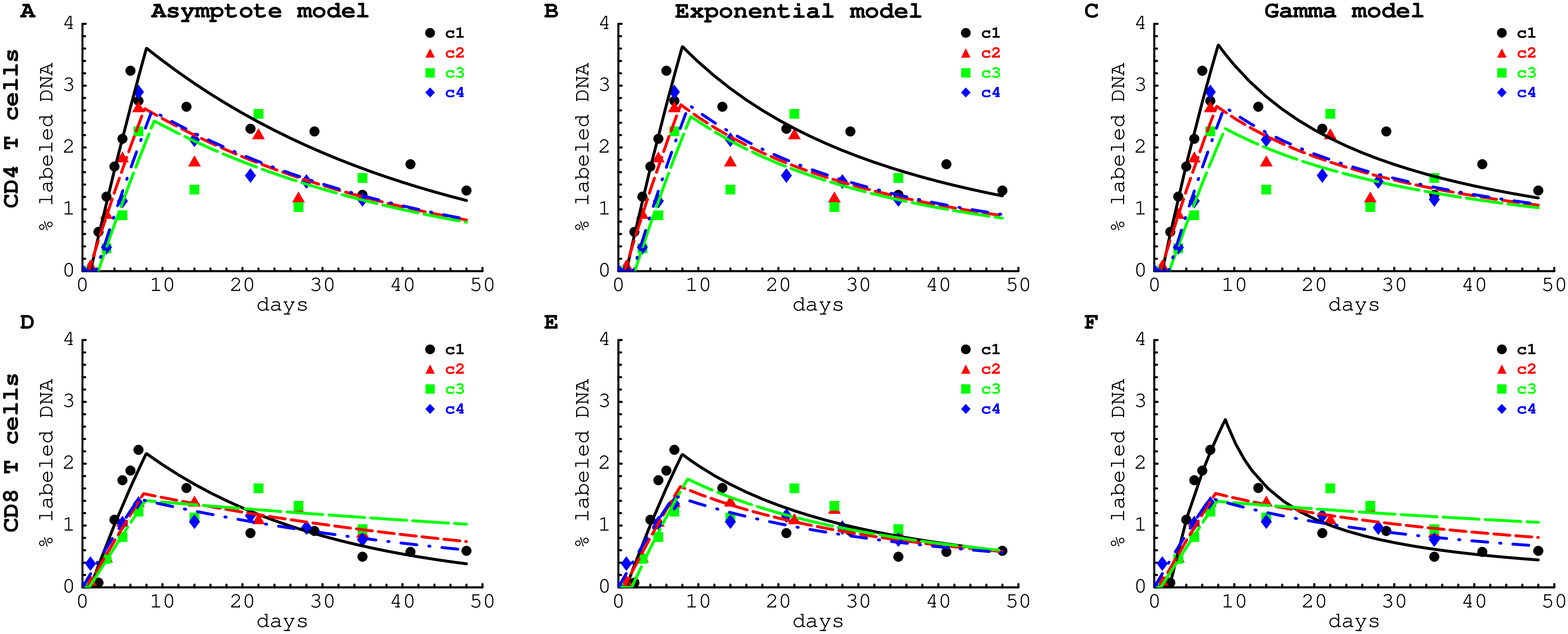} 
\ec
\caption{Fitting the deuterium labeling data of CD4$^+$ (top rows) and
  CD8$^+$ (bottom rows) T cells in four healthy humans with three
  models: the Asymptote model (panels A and D), the Exponential model,
  in which a fraction $\al$ of cells have exponentially-distributed
  turnover rates (\eref{L-expa}, panels B and E), and the Gamma model,
  with gamma-distributed turnover rates (\eref{L-gamma}, panels C and
  F). Experimental data obtained from \citet{Mohri.jem01} are shown as
  symbols and the curves are the best model fits. The sum of squared
  residuals of the model fits to the data on the dynamics of CD4$^+$ T
  cells are $(6.19,5.94,5.87)\times10^{-3}$ for the Asymptote model,
  the Exponential model and the Gamma model, respectively. The sum of
  squared residuals of the model fits to the data on the dynamics of
  CD8$^+$ T cells are $(3.4,3.85,1.56)\times10^{-3}$ for the Asymptote
  model, the Exponential model and the Gamma model, respectively. Note
  that the two explicit kinetic heterogeneity models describe these
  data with similar (Exponential model) or even better (Gamma model)
  quality compared to the Asymptote model. }
    \label{fig:mohri.controls}
\end{figure}

\begin{figure}
\bc \includegraphics[width=0.99\textwidth]{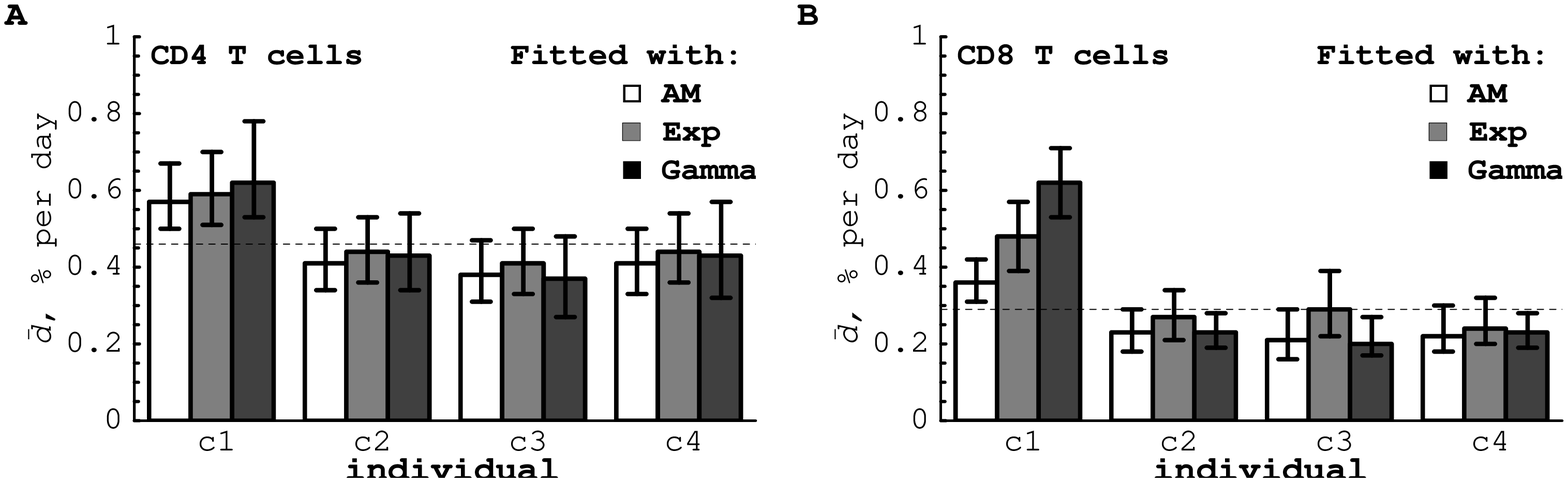} \ec
\caption{Estimates of the average turnover rates of CD4$^+$ (panel A)
  and CD8$^+$ (panel B) T cells in four healthy humans obtained by
  fitting with three different models: the Asymptote model ($\Box$,
  the Exponential model (\textcolor{light-gray}{$\blacksquare$}), in
  which a fraction $\al$ of the cells have exponentially-distributed
  turnover rates, and the Gamma model
  (\textcolor{dark-gray}{$\blacksquare$}) with gamma-distributed
  turnover rates. Best fits of the data are shown in
  \fref{mohri.controls}, and estimates of all parameters of the models
  are shown in Appendix (\tref{mohri.controls.cd4} and
  \ref{tab:mohri.controls.cd8}).  Confidence intervals were obtained
  by bootstrapping the residuals with 1000 simulations. Horizontal
  dashed lines denote the mean of the estimated turnover rates in all
  patients and all models for CD4$^+$ ($\pp = 0.46\%$ per day) and
  CD8$^+$ ($\pp = 0.29\%$ per day) T cells.  Note that all models
  deliver very similar estimates for the average turnover rate $\pp$,
  with the exception of the estimated CD8$^+$ T-cell turnover rates in
  individual c1 which are highly model-dependent.}
\label{fig:mohri.controls.p}
\end{figure}

\section{Discussion}

In this paper we have analyzed the models that are commonly used in
the literature to estimate the rates of cell turnover from deuterium
labeling data. We have shown that the three most commonly used models
are mathematically identical and therefore provide identical fits to
the data, and only differ in the biological interpretation of the
estimated parameters \cite{Asquith.ti09}. The simplest summary of
labeling data is provided by a model that has two parameters: $d$ as
the rate of cell death in the population, and $\al$ as the fraction of
cells that undergo turnover, which determines the asymptote of the
uplabeling phase (see \eref{L-deboer}). In this model, $\al\times d$
gives the estimated average rate of cell turnover. We have extended
this model by allowing for multiple sub-populations $i$ of size
$\al_i$ with different turnover rates $d_i$ (see \eref{L-general}).
This extended model can be used to investigate potential heterogeneity
of cell populations, by fitting labeling data with a model that has
one, two, or more sub-populations with different turnover rates. Using
standard techniques of model selection (e.g., the partial F-test or
the Akaike Information Criterion), one can investigate which of those
models describes the labeling data best, given the number of model
parameters \cite{Bates.b88,Burnham.b02}, or one can study whether the
estimated average turnover rate is converging to an invariant value by
increasing the number of compartments (work in progress).

For the case where the number of sub-populations is large, we derived
a model with continuous kinetic heterogeneity. For several continuous
distributions such as the exponential and the gamma distribution, the
model predicts that the initial rate of loss of labeled DNA after
label withdrawal is determined by the duration of the labeling period
as has been observed experimentally \cite{Asquith.ti02}. Moreover, in
the model the average turnover rate, which determines the initial rate
of label accumulation in the population, turned out to be independent
of the length of the labeling period. However, it should be noted that
the average rate of cell turnover that is estimated from experimental
data using, for example, the Asymptote model, may in fact depend on
the duration of labeling \cite[Den Braber et al. (in
prepartion)]{Borghans.ir07}. Potential reasons for this discrepancy
will be investigated in more detail elsewhere.

Previous models had certain artifacts: the asymptote labeling level
was dependent on the length of the labeling period, and the accrual of
labeled DNA during the uplabeling phase and the loss of labeled DNA
during the downlabeling period were always described by single
exponential functions. It is, therefore, unclear whether such limited
models provide a good description of truly kinetically heterogeneous
populations.  Although at first glance it may seem impossible to fit
labeling data with models that contain an infinite number of
sub-populations with different rates of turnover, we have shown that
this is in fact feasible. By fitting artificial labeling data, we have
validated these new models: they generally give good fits to the data
and converge on average turnover rates that are close to the known
average turnover rate.  Moreover, the new explicit heterogeneity model
outperformed the Asymptote model when it came to fitting experimental
data, especially when the rates of label accumulation and loss are not
exponential (see \fref{mohri.controls}). Importantly, due to its
relatively general structure, all results of the kinetic heterogeneity
model are applicable to both non-proliferating and proliferating
lymphocytes, all having a distribution of turnover values (results not
shown). Moreover, because the model naturally incorporates the
dependence of the rate of label loss on the length of the labeling
period, this is the first model that can be strictly applied to fit
labeling data with different labeling periods.


We have focused our analysis on a particular type of kinetic
heterogeneity in which kinetic properties of cells of a given
subpopulation do not change over time and there is no exchange of
cells in different subpopulations. However, in some circumstances,
such as during an immune response, this assumption maybe violated
since over the course of infection, lymphocytes do changes their
kinetic properties over time (e.g., \cite{Antia.nri05}). Taking into
account such a type of \textit{temporal} heterogeneity is therefore
biologically justified.  We have shown that estimates of the average
turnover rate of the population may depend on the model used (e.g.,
\fref{simdata} and \ref{fig:mohri.controls.p}B); it is therefore
important to develop additional (simple) models that take temporal
heterogeneity of cell populations into account
\cite{Grossman.s99,Grossman.nm06}. We propose that future studies
should aim at testing multiple models in how well they describe the
labeling data and whether these models deliver similar estimates of
important kinetic parameters such as the average rate of cell
turnover.

\section{Methods}

When fitting experimental data, the models were extended to allow for
the initial delay in the labeling of cells (see also
\cite{Mohri.jem01}). For example, including a delay in the Asymptote
model (given by \eref{L-deboer}) takes the form

\beq L(t)=\left\{\begin{array}{ll}
    0, & \mbox{if $t\leq\tau$},\\
    \al(1-e^{-d(t-\tau)}), &
    \mbox{if $\tau<t\leq T+\tau$},\\
    L(T)e^{-d(t-T-\tau)}, & \mbox{otherwise}.\\
\end{array} \right. \label{eqn:L-deboer-tau}
\ee

To normalize the residuals of the model fits to experimental data,
given that the data are expressed as proportions, the data and the
model predictions were transformed as $\arcsin(\sqrt x)$ where $x$ is
the frequency of labeled DNA in the population \cite{Hogg.b95}. The
models were fitted according to the least squares method by using the
FindMinimum routine in Mathematica. Confidence intervals were
calculated by bootstrapping the residuals with 1000 simulations.

\bibliography{/home/fly10/vitaly/refs/bibliography/library-main}

\begin{thebibliography}{23}
\expandafter\ifx\csname natexlab\endcsname\relax\def\natexlab#1{#1}\fi

\bibitem[Hellerstein et~al.(1999)]{Hellerstein.nm99}
Hellerstein, M., M.~B. Hanley, D.~Cesar, S.~Siler, C.~Papageorgopoulos,
  E.~Wieder, D.~Schmidt, R.~Hoh, R.~Neese, D.~Macallan, S.~Deeks, and J.~M.
  McCune. 1999. Directly measured kinetics of circulating {T} lymphocytes in
  normal and {HIV}-1-infected humans {\em Nat Med\/} 5: 83--9.

\bibitem[Mohri et~al.(2001)]{Mohri.jem01}
Mohri, H., A.S. Perelson, K.~Tung, R.M. Ribeiro, B.~Ramratnam, M.~Markowitz,
  R.~Kost, A.~Hurley, L.~Weinberger, D.~Cesar, M.K. Hellerstein, and D.D. Ho.
  2001. Increased turnover of {T} lymphocytes in {HIV}-1 infection and its
  reduction by antiretroviral therapy. {\em J Exp Med\/} 194: 1277--87.

\bibitem[Vrisekoop et~al.(2008)]{Vrisekoop.pnas08}
Vrisekoop, N., I.~den Braber, A.~B. de~Boer, A.~F. Ruiter, M.~T. Ackermans,
  S.~N. van~der Crabben, E.~H. Schrijver, G.~Spierenburg, H.~P. Sauerwein,
  M.~D. Hazenberg, R.~J. de~Boer, F.~Miedema, J.~A. Borghans, and K.~Tesselaar.
  2008. Sparse production but preferential incorporation of recently produced
  naive {T} cells in the human peripheral pool. {\em Proc Natl Acad Sci U S
  A\/} 105: 6115--20.

\bibitem[Hellerstein(1999)]{Hellerstein.it99}
Hellerstein, M.K. 1999. Measurement of {T}-cell kinetics: recent methodologic
  advances. {\em Immunol Today\/} 20: 438--41.

\bibitem[Macallan et~al.(1998)]{Macallan.pnas98}
Macallan, D.~C., C.~A. Fullerton, R.~A. Neese, K.~Haddock, S.~S. Park, and
  M.~K. Hellerstein. 1998. Measurement of cell proliferation by labeling of
  {DNA} with stable isotope-labeled glucose: studies in vitro, in animals, and
  in humans {\em Proc Natl Acad Sci USA\/} 95: 708--13.

\bibitem[Neese et~al.(2001)]{Neese.ab01}
Neese, R.A., S.Q. Siler, D.~Cesar, F.~Antelo, D.~Lee, L.~Misell, K.~Patel,
  S.~Tehrani, P.~Shah, and M.K. Hellerstein. 2001. Advances in the stable
  isotope-mass spectrometric measurement of {DNA} synthesis and cell
  proliferation. {\em Anal Biochem\/} 298: 189--95.

\bibitem[Borghans and de~Boer(2007)]{Borghans.ir07}
Borghans, J.A. and R.J. de~Boer. 2007. Quantification of {T}-cell dynamics:
  from telomeres to {DNA} labeling. {\em Immunol Rev\/} 216: 35--47.

\bibitem[Ribeiro et~al.(2002{\natexlab{a}})]{Ribeiro.pnas02}
Ribeiro, R.M., H.~Mohri, D.D. Ho, and A.S. Perelson. 2002{\natexlab{a}}. In
  vivo dynamics of {T} cell activation, proliferation, and death in {HIV}-1
  infection: why are {CD}4+ but not {CD}8+ {T} cells depleted? {\em Proc Natl
  Acad Sci USA\/} 99: 15572--7.

\bibitem[Ribeiro et~al.(2002{\natexlab{b}})]{Ribeiro.bmb02}
Ribeiro, R.M., H.~Mohri, D.D. Ho, and A.S. Perelson. 2002{\natexlab{b}}.
  Modeling deuterated glucose labeling of {T}-lymphocytes. {\em Bull Math
  Biol\/} 64: 385--405.

\bibitem[Asquith et~al.(2002)]{Asquith.ti02}
Asquith, B., C.~Debacq, D.C. Macallan, L.~Willems, and C.R. Bangham. 2002.
  Lymphocyte kinetics: the interpretation of labelling data. {\em Trends
  Immunol\/} 23: 596--601.

\bibitem[De~Boer et~al.(2003)]{DeBoer.pbs03}
De~Boer, R.J., H.~Mohri, D.D. Ho, and A.S. Perelson. 2003. Estimating average
  cellular turnover from 5-bromo-2'-deoxyuridine ({B}rd{U}) measurements. {\em
  Proc R Soc Lond B Biol Sci\/} 270: 849--58.

\bibitem[Asquith et~al.(2009)]{Asquith.ti09}
Asquith, B., J.A. Borghans, V.V. Ganusov, and D.C. Macallan. 2009. Lymphocyte
  kinetics in health and disease. {\em Trends Immunol\/} .

\bibitem[Parretta et~al.(2008)]{Parretta.ji08}
Parretta, E., G.~Cassese, A.~Santoni, J.~Guardiola, A.~Vecchio, and F.~Di~Rosa.
  2008. Kinetics of in vivo proliferation and death of memory and naive {CD}8
  {T} cells: parameter estimation based on 5-bromo-2'-deoxyuridine
  incorporation in spleen, lymph nodes, and bone marrow. {\em J Immunol\/} 180:
  7230--9.

\bibitem[Bates and Watts(1988)]{Bates.b88}
Bates, D.~M. and D.~G. Watts. 1988. {\em Nonlinear regression analysis and its
  applications.\/} John Wiles \& Sons, Inc. 365 .

\bibitem[Davenport et~al.(2004)]{Davenport.jv04}
Davenport, M.P., R.M. Ribeiro, and A.S. Perelson. 2004. Kinetics of
  virus-specific {CD}8+ {T} cells and the control of human immunodeficiency
  virus infection. {\em J Virol\/} 78: 10096--103.

\bibitem[Ganusov and De~Boer(2007)]{Ganusov.ti07}
Ganusov, V.V. and R.J. De~Boer. 2007. Do most lymphocytes in humans really
  reside in the gut? {\em Trends Immunol\/} 28: 514--8.

\bibitem[Murali-Krishna et~al.(1998)]{Murali-Krishna.i98}
Murali-Krishna, K., J.D. Altman, M.~Suresh, D.J.D. Sourdive, A.J. Zajac, J.D.
  Miller, J.~Slansky, and R.~Ahmed. 1998. Counting antigen-specific {CD}8+ {T}
  cells: {A} re-evaluation of bystander actiation during viral infection {\em
  Immunity\/} 8: 177--187.

\bibitem[Homann et~al.(2001)]{Homann.nm01}
Homann, D., L.~Teyton, and M.B. Oldstone. 2001. Differential regulation of
  antiviral {T}-cell immunity results in stable {CD}8+ but declining {CD}4+
  {T}-cell memory. {\em Nat Med\/} 7: 913--919.

\bibitem[Burnham and Anderson(2002)]{Burnham.b02}
Burnham, K.~P. and D.~R. Anderson. 2002. {\em Model selection and multimodel
  inference: a practical information-theoretic approach.\/} Springer-Verlag,
  New York 340 .

\bibitem[Antia et~al.(2005)]{Antia.nri05}
Antia, R., V.V. Ganusov, and R.~Ahmed. 2005. The role of models in
  understanding {CD}8+ {T}-cell memory. {\em Nat Rev Immunol\/} 5: 101--111.

\bibitem[Grossman et~al.(1999)]{Grossman.s99}
Grossman, Z., R.~B. Herberman, D.~S. Dimitrov, I.~M. Rouzine, and J.~M. Coffin.
  1999. T cell turnover in {SIV} infection {\em Science\/} 284: 555a--555d.

\bibitem[Grossman et~al.(2006)]{Grossman.nm06}
Grossman, Z., M.~Meier-Schellersheim, W.E. Paul, and L.J. Picker. 2006.
  Pathogenesis of {HIV} infection: what the virus spares is as important as
  what it destroys. {\em Nat Med\/} 12: 289--95.

\bibitem[Hogg and Craig(1995)]{Hogg.b95}
Hogg, R.V. and A.T. Craig. 1995. {\em Introduction to {M}athematical
  {S}tatistics\/} Macmillan.

\end{thebibliography}

\newpage
\section{Supplementary Information}

\setcounter{equation}{0}
\renewcommand{\theequation}{S.\arabic{equation}}

\subsection{Derivation of the model with continuous kinetic
  heterogeneity}

The average label incorporation in a population consisting of $n$
sub-populations is given by \eref{L-general}. If the number of
sub-populations is large ($n\ra\infty$), one could switch from
summation to integration in \eref{L-general}. If $\al_i$ is the
fraction of cells in sub-population $i$ with turnover rate $d_i$,
then, as $n\ra\infty$, $\al_i = \ff(d)\id d$, where the latter is the
probability that a randomly chosen cell from a population will have a
turnover rate in the range $(d,d+\id d)$. If $\sum_{i=1}^n \al_i = 1$,
then $\ff(d)$ is also normalized to 1.  Then for $t>T$,
\eref{L-general} can be rewritten in a different form

\beq L(t)=\lim_{n\ra\infty}\left[ 1- \sum_{i=1}^n \al_i {\rm e}^{-d_i
    t}\right] = 1-\int_0^\infty \ff(d){\rm e}^{-d t}\id
d\label{eqn:L-pulse2}.\ee

Similarly, from \eref{L-general}, during the delabeling period, the
fraction of labeled DNA is given by

\beq L(t)=\int_0^\infty \ff(d){\rm e}^{-d (t-T)}\id d -\int_0^\infty
\ff(d){\rm e}^{-d t}\id d.
\label{eqn:L-chase2} \ee

One can then calculate several important characteristics
that determine the change of the fraction of labeled DNA over time. First, the
initial uplabeling rate can be calculated using \eref{L-pulse2} for
small $t$, yielding

\beq L(t) = 1-\int_0^\infty \ff(d){\rm e}^{-d t}\id
d\approx 1-\int_0^\infty \ff(d)(1-d t)\id
d = \pp t\label{eqn:init_rate}.\ee

\no where $\pp = \int_0^\infty d\ff(d)\id d$, and $\int_0^\infty
\ff(d)\id d=1$ by definition. The importance of this result is that it
demonstrates that for any distribution $\ff(d)$, the estimated initial
rate of uplabeling is determined only by the average rate of cell
turnover.

During delabeling, the initial change in the fraction of labeled DNA (for
$t=T+\eps$ with $\eps$ relatively small), can be calculated from:

\beqa L(t)&=&\int_0^\infty \ff(d){\rm e}^{-d (t-T)}\id d
-\int_0^\infty \ff(d){\rm e}^{-d t}\id d = \int_0^\infty
\ff(d)(1-d\eps)\id
d -\nonumber\\
& &\int_0^\infty \ff(d){\rm e}^{-d T}(1-d\eps)\id d = 1-\int_0^\infty
\ff(d){\rm e}^{-d T}\id d - \eps\left[\pp -\int_0^\infty d\ff(d){\rm
    e}^{-d T}\id d\right] = \nonumber\\
&&L(T)-\eps L(T)\left[{\pp - \int_0^\infty d\ff(d){\rm e}^{-d T}\id
    d\over 1-\int_0^\infty \ff(d){\rm e}^{-d T}\id d}\right],
\label{eqn:L-chase3} \eea

\no where $L(T)=1-\int_0^\infty \ff(d){\rm e}^{-d T}\id d$. Then the
initial per capita rate of loss of labeled DNA, $d^*$, can be
calculated for short and long labeling periods. For short labeling
periods $T\ra0$, and we find

\beqa d^*&=&{\pp - \int_0^\infty d\ff(d){\rm e}^{-d T}\id d\over
  1-\int_0^\infty \ff(d){\rm e}^{-d T}\id d}= {\pp - \int_0^\infty
  d\ff(d)(1-d T)\id d\over 1-\int_0^\infty \ff(d)(1-d T)\id
  d}={\overline{d^2}\over \pp}=\pp + {\mbox{var}(d)\over \pp},
\label{eqn:d1} \eea

\no where $\mbox{var}(d)=\overline{d^2}-(\pp)^2$ is the variance of
the turnover rates in the population. When the labeling period is long,
$T\ra\infty$, the per capita rate of label loss is

\beqa d^*&=&{\pp - \int_0^\infty d\ff(d){\rm e}^{-d T}\id d\over
  1-\int_0^\infty \ff(d){\rm e}^{-d T}\id d}= \pp,
\label{eqn:d2} \eea

\no since all terms ${\rm e}^{-dT}\ra0$ as $T\ra\infty$. This confirms
the conjecture of \citet{Asquith.ti02} that $d^*$ should approach the
average rate of turnover after long labeling period. We now add that
the maximal difference between the average turnover rate, $\pp$, and
the loss rate of labeled cells, $d^*$, is set by the distribution of
turnover rates in the population.

\subsection{Particular solutions of the kinetic heterogeneity model}

For several simple distributions of the turnover rates, we can
obtain analytical solutions for the change in the fraction of labeled
DNA during the labeling experiment.

{\bf Exponential distribution}. In this case $\ff(d)=(1/\pp) {\rm
  e}^{-d/\pp}$, where $\pp$ is the average turnover rate in the
population. Using \eref{L-pulse2} and (\ref{eqn:L-chase2}), we find

\beqa L(t)&=& 1-{1\over \pp}\int_0^\infty {\rm e}^{-d t-d/\pp}\id
d=1-{1\over (t+1/\pp)\pp}={\pp t\over 1+\pp t},\quad t\leq T,
\label{eqn:L-pulse3}\\
L(t)&=& {1\over \pp}\int_0^\infty {\rm e}^{-d (t-T)-d/\pp}\id d -
{1\over \pp}\int_0^\infty {\rm e}^{-d t-d/\pp}\id d={1\over 1+
  \pp(t-T)}-{1\over 1+\pp t}\nonumber\\
&=& L(T){1+\pp T\over (1+\pp t)(1+\pp(t-T))} ,\quad
t>T.\label{eqn:L-chase31}\eea

When $\pp t\ll 1$, the fraction of labeled DNA is simply $L(t)\approx
\pp t$, i.e., the initial rate of increase is again given by the
average turnover rate $\pp$. The initial rate of decline during
delabeling is less clear. Let us define $t=T+\eps$ where $\bar p
\eps\ll 1$. Then after cessation of label administration, using
Taylor's expansion we find

\beqa L(t)&=& {1\over 1+\pp\eps}-{1\over 1+\pp T+\pp\eps}=
1-\pp \eps-{1\over 1+\pp T}+ {\pp\eps\over (1+\pp
  T)^2}+o(\eps)=\nonumber \\
&=&L(T)-L(T)\pp \left(1+{1\over 1+\pp
    T}\right)\eps+o(\eps)\label{eqn:L-chase4} \eea

\no where $L(T)=1-1/(1+\pp T)$. This expression shows that the initial
per capita rate of loss of labeled DNA, $d^*= \pp \left(1+{1\over
    1+\pp T}\right)$, decreases with increasing length of the labeling
period \cite{Asquith.ti02}.  If $\pp T\ll 1$ (short labeling), the
initial per capita decay rate is $d^*\approx 2\pp$.

{\bf Gamma distribution}. In this case $\ff(d)=\lam (\lam
d)^{k-1}{\rm e}^{-\lam d}/(k-1)!$, where $\lam$ and $k$ are the scale
and shape parameters, respectively, and the average rate of cell
turnover $\pp=k/\lam$, $\sigma^2_d=\pp^2/k$, and $CV=\sigma_d/\pp =
1/\sqrt{k}$. Note that when $k=1$, the gamma distribution is identical to
the exponential distribution. Substituting the gamma distribution in
\eref{L-pulse2}, we find

\beqa L(t)&=& 1-{\lam^{k}\over (k-1)!}\int_0^\infty d^{k-1}{\rm e}^{-d
  t-\lam d}\id d\nonumber \\&=& 1-{\lam^{k}\over
  (k-1)!(t+\lam)^{k}}\int_0^\infty d^{k-1}(t+\lam)^{k}{\rm
  e}^{-d(t+\lam)}\id d \nonumber \\
& = & 1-{\lam^{k}\over (t+\lam)^k}=1-\left[1+ {\pp t\over
    k}\right]^{-k},\quad t\leq T.\label{eqn:L-pulse5}
\eea

Using \eref{L-chase2} and proceeding similarly, we find the change in
the fraction of labeled DNA during delabeling:

\beqa L(t)&=& \left[1+{\pp(t-T)\over k}\right]^{-k}-\left[1+  {\pp t\over
    k}\right]^{-k},\quad t> T.\label{eqn:L-chase5}\\
\eea

The initial rate of increase in the fraction of labeled cells is also
independent of the length of the labeling period and, initially, for
$t=\eps$ (such as $\pp\eps\ll1$) is

\beqa L(\eps)=1-\left[{1\over 1+ {\pp\eps\over
      k}}\right]^{k}=1-(1-k\pp\eps/k)+o(\eps)=\pp\eps+o(\eps).
\label{eqn:L-gamma2} \eea

\no as expected. The initial per capita rate of loss of labeled DNA is
somewhat more complex. For times $t=T+\eps$ such as $\pp\eps\ll 1$,
using Taylor's expansion, we find

\beqa L(t)&=& \left[1+{\pp\eps\over k}\right]^{-k}-\left[1+
  {\pp(T+\eps)
\over k}\right]^{-k}=1-{1\over \left(1+{\pp T\over
      k}\right)^{k}}-\pp\eps \left[1-{1\over \left(1+{\pp T\over
      k}\right)^{k}}\right].\label{eqn:L-gamma3} \eea

It is useful to rewrite this expression in terms of $L(T)$:

\beqa L(t)&=& L(T)-L(T){\pp\eps\over 1+\pp T/k} \left[{(1+\pp 
T/k)^{k+1}-1\over (1+\pp T/k)^{k}-1}\right] =
L(T)(1-d^*\eps).\label{eqn:L-gamma4} \eea

\no where $d^* = {\pp\over 1+\pp T/k} \left[{(1+\pp 
T/k)^{k+1}-1\over (1+\pp T/k)^{k}-1}\right]$ is the initial per
capita loss of labeled  DNA. For short labeling ($\pp T\ll 1$), the
initial per capita decay rate is $d^*\approx \pp(k+1)/k$, and as the
shape parameter $k$ becomes larger, the decline rate $d^*$ approaches
the average proliferation rate $\pp$. For long labeling periods ($\pp T\gg
1$), $d^*\approx \pp$, as expected.

{\bf Truncated gamma distribution}. Under some circumstances the
distribution of turnover rates may allow for too high rates of cell
turnover. To circumvent this problem one may use a truncated
distribution. For a gamma distribution truncated at maximal value
$\pmax$, the distribution is similar as above with an added
normalization constant $C$, $\ff(d) = C^{-1} \lam (\lam d)^{k-1}{\rm
  e}^{-\lam d}/(k-1)!$. The constant $C$ is found by normalizing the
probability distribution

\beq C = \int_0^{\pmax} \ff(d)\id d = 1 - {\Gamma(k,\pmax\lam)\over
  (k-1)!},
\label{eqn:constant}
\ee

\no where $\Gamma(k,d)=\int_{d}^\infty x^{k-1} e^{-x} \id x$ is an 
incomplete gamma function. The average turnover rate then has to be
calculated numerically

\beq \pp = C^{-1} \int_0^{\pmax} d \ff(d)\id d = {k\over\lam}\left[ 
{ 1  - {\Gamma(k+1,\pmax\lam)\over k!} \over 1 - {\Gamma(k,\pmax\lam)\over
    (k-1)! }}  \right].
\label{eqn:p-average}
\ee

For the fraction of labeled nucleotides, we proceed as in
\eref{L-pulse5} and obtain

\beqa L(t)&=& 1-{\lam^{k}\over C (k-1)!}\int_0^{\pmax} d^{k-1}{\rm
  e}^{-d t-\lam d}\id d\nonumber \\&=& 1-{\lam^{k}\over C
  (k-1)!(t+\lam)^{k}}\int_0^{\pmax(t+\lam)} x^{k-1}{\rm
  e}^{-x}\id x \nonumber \\
& = & 1-{\lam^{k}\over C (k-1)!(t+\lam)^{k}}\left[(k-1)! -
  \int_{\pmax(t+\lam)}^\infty x^{k-1}{\rm
    e}^{-x}\id x\right] \nonumber \\
& = & 1-{\lam^{k}\over (t+\lam)^k} \times
{(k-1)! - \Gamma(k,\pmax(\lam+t))\over (k-1)!- \Gamma(k,\pmax\lam) },\quad
t\leq T.\label{eqn:L-gamma5}
\eea

During delabeling ($t>T$), we proceed as in \eref{L-chase5}
and find

\beqa L(t)&=& {\lam^{k}\over (\lam+t-T)^k}\times {(k-1)! -
  \Gamma(k,\pmax(\lam+t-T))\over (k-1)!- \Gamma(k,\pmax\lam) } -\nonumber\\
&& {\lam^{k}\over (t+\lam)^k} \times {(k-1)! -
  \Gamma(k,\pmax(\lam+t))\over (k-1)!- \Gamma(k,\pmax\lam) }
.\label{eqn:L-gamma6}
\eea

Since $\lim_{\pmax\ra\infty}\Gamma(k,\pmax)=0$, at $\pmax\ra\infty$,

the fraction of labeled nucleotides becomes identical to
\eref{L-gamma} with $\lam = k/\pp$.

\subsection{Non-parametric estimates of the distribution of
  proliferation rates in the population}

Mathematically, the last term in \eref{L-pulse2} is a Laplace transformation
of $\ff(p)$. This result stems from the assumption of exponentially
distributed inter-division times of cells and raises the intriguing
possibility that from the change in the fraction of labeled DNA during
label administration, one can estimate the distribution of
proliferation rates in the population. Let us denote $\ff^*(t)$
as the Laplace transformation of $\ff(d)$:

\beq \ff^*(t) =\mathcal{L}[\ff(d)]= \int_0^\infty \ff(d){\rm
  e}^{-d t}\id d.
\label{eqn:zeta-1} \ee

From \eref{L-pulse2}, one finds the distribution of proliferation
rates in the population using the inverse Laplace transformation:

\beq \ff(d) =\mathcal{L}^{-1}[1-L(t)].
\label{eqn:zeta-2} \ee

\no where $L(t)$ is a curve describing the change of the fraction of
labeled DNA during label administration. Such a curve could be
obtained in several ways, for example, by interpolating the data.
Application of this method for the analysis of experimental data will
be published elsewhere.

\begin{table}
\renewcommand{\arraystretch}{1.5}
\small 
\begin{tabular}{|cc|c|c|c|c|c|c|c|}
  \hline
  \multirow{15}{*}{\rotatebox{90}{\bf Data fitted with:}}&\multicolumn{8}{c|}
  {\large \bf Data generated using:}\\

&\multicolumn{2}{c}{}&\multicolumn{2}{c}{Gamma model}&
\multicolumn{2}{c}{Exponential model}&
\multicolumn{2}{c|}{Two-populations model}\\ \cline{4-9}
&\multicolumn{2}{c|}{}&7 days&15 days&7 days&15 days&7 days&15 days\\ \cline{3-9}
&\multirow{4}{*}{\rotatebox{90}{Asymptote}}
&$\pp$&0.076&0.068&0.082&0.08&0.04&0.039\\
&&&0.068---0.085&0.06---0.077&0.073---0.096&0.07---0.092&0.034---0.046&0.032---0.047\\
&&$d$&0.121&0.117&0.203&0.199&0.052&0.054\\
&&&0.105---0.14&0.099---0.137&0.17---0.238&0.169---0.231&0.04---0.066&0.038---0.071\\
\cline{3-9}
&\multirow{4}{*}{\rotatebox{90}{Exponential}}&$\pp$&0.09&0.09&0.11&0.12&0.05&0.05\\
&&&0.082---0.104&0.076---0.1&0.105---0.122&0.104---0.137&0.042---0.055&0.040---0.055\\
&&$\alpha$&0.83&0.76&0.51&0.48&1&1\\
&&&0.773---0.908&0.70---0.82&0.488---0.536&0.456---0.51&0.857---1.0&0.835--1.0\\
\cline{3-9}
&\multirow{4}{*}{\rotatebox{90}{Gamma}}&$\pp$&0.097&0.099&0.154&0.197&0.044&0.044\\
&&&0.085---0.112&0.085---0.117&0.14---0.169&0.152---0.271&0.038---0.052&0.036---0.054\\
&&$k$&0.594&0.44&0.197&0.159&1.329&1.132\\
&&&0.478---0.768&0.367---0.536&0.184---0.211&0.134---0.183&0.796---2.947&0.679---2.901\\
\hline
\end{tabular}
\caption{Estimates of the parameters after fitting three models to
  three sets of artificial data. Rows correspond to different models
  used to fit the data, and columns correspond to the models used to
  generate the data.  We show parameter estimates of the Asymptote
  model, the Exponential model (in which a fraction $\al$ of cells
  have exponentially distributed turnover rates), and the Gamma model
  (with gamma distributed turnover rates). Data were generated using
  the Gamma model (\eref{L-gamma}), the Exponential model
  (\eref{L-expa}), and the Two-populations model (\eref{L-general}).
  The 95\% confidence intervals, that are shown below the mean values,
  were obtained
  by bootstrapping the residuals with 1000 simulations. Data have been
  generated for two labeling periods, of 7 and 15 days, respectively.
  For all data, the average rate of turnover was fixed at $0.1$/day.
  Other parameters used to generate the data are: $k=0.5$ (Gamma
  model), $\al=0.5$ and $\ppa=0.2$/day (Exponential model),
  $d_1=1$/day, and $\al=0.07$ (Two-populations model).
}\label{tab:simdata}
\end{table}

\begin{table}
\renewcommand{\arraystretch}{1.5}
\small 
\bc
\begin{tabular}{|l|l|l|l|}
\hline
&\multicolumn{3}{c|} {\large \bf Data fitted with:}\\
   &\multicolumn{1}{|c}{Asymptote model}&\multicolumn{1}{c}{Exponential model}&Gamma model\\ \hline
  $d/\ppa/k$&2.87 (2.22---3.65)&1.94 (1.46---2.6)&0.12 (0.09---0.17)\\
  \hline
  $\pp_1$, \% $day^{-1}$&0.57 (0.5---0.67)&0.59 (0.51---0.7)&0.62(0.53---0.78)\\
  $\pp_2$&0.41 (0.34---0.5)&0.44 (0.36---0.53)&0.43 (0.34---0.54)\\
  $\pp_3$&0.38 (0.31---0.47)&0.41 (0.33---0.5)&0.37 (0.27---0.48)\\
  $\pp_4$&0.41 (0.33---0.5)&0.44 (0.36---0.54)&0.43 (0.32---0.57)\\
  \hline
  $\tau_1$, $day$&1. (0.91---1.51)&1. (0.93---1.53)&1. (0.93---1.57)\\
  $\tau_2$&0.78 (0.28---1.)&0.81 (0.35---1.)&0.8 (0.31---1.)\\
  $\tau_3$&1.97 (1.---2.65)&2.06 (1.14---2.7)&1.87 (0.73---2.67)\\
  $\tau_4$&1.7 (0.98---2.45)&1.83 (1.---2.54)&1.79 (1.---2.58)\\
  \hline
  RSS, $10^{-3}$&6.19&5.94&5.87\\
  \hline
\end{tabular}
\end{center}
\caption[]{Average turnover rates of CD4$^+$ T cells from four healthy
  humans as estimated by fitting the data from \citet{Mohri.jem01}
  using the Asymptote model, the Exponential model, and the Gamma
  model. The best fits of the models resulted in different average
  rates of cell turnover $\pp_i$ and initial delays of labeling
  $\tau_i$. Other parameters, that could be assumed to be identical
  between different patients, are the death rate of labeled cells $d$
  (Asymptote model), the rate of turnover $\ppa$ of the turning-over
  sub-population in the Exponential model, and the shape parameter $k$
  in the model with gamma distributed turnover rates. For the model
  with gamma distributed turnover rates, an asymptote level $\al =1$
  provided the best fit of the data. The quality of the fit is
  illustrated by the residual sum of squares (RSS).  The 95\%
  confidence intervals were obtained by bootstrapping the residuals
  with 1000 simulations.}
\label{tab:mohri.controls.cd4}
\end{table}

\begin{table}
\renewcommand{\arraystretch}{1.5}
\small 
\bc
\begin{tabular}{|l|l|l|l|}
\hline
&\multicolumn{3}{c|} {\large \bf Data fitted with:}\\
   &\multicolumn{1}{|c}{Asymptote
     model}&\multicolumn{1}{c}{Exponential model}&Gamma model\\ \hline
$\al_1/\al/k_1$&0.08 (0.07---0.10)&0.13 (0.12---0.16)&0.033 (0.028---0.039)\\
  $\al_2/\al/k_2$&0.13 (0.09---0.73)&0.13 (0.12---0.16)&0.116 (0.066---0.341)\\
  $\al_3/\al/k_3$&0.26 (0.10---1.0)&0.13 (0.12---0.16)&0.347 (0.099---$10^7$)\\
  $\al_4/\al/k_4$&0.10 (0.07---0.28)&0.13 (0.12---0.16)&0.082 (0.05---0.155)\\
  \hline
  $\pp_1$, 
\% per day&0.36 (0.31---0.43)&0.48 (0.39---0.57)&0.62 (0.53---0.71)\\
   $\pp_2$&0.23 (0.18---0.30)&0.27 (0.21---0.34)&0.23 (0.19---0.28)\\
   $\pp_3$&0.21 (0.17---0.31)&0.29 (0.22---0.39)&0.2 (0.17---0.27)\\
   $\pp_4$&0.22 (0.18---0.3)&0.24 (0.2---0.32)&0.23 (0.19---0.28)\\
  \hline
  $\tau_1$&0.99 (0.73---1.38)&1.76 (1.33---1.95)&1.89 (1.75---1.98)\\
  $\tau_2$&0.65 (0.---0.94)&0.76 (0.11---0.99)&0.65 (0.28---0.92)\\
  $\tau_3$&0.85 (0.---1.99)&1.73 (0.59---2.4)&0.82 (0.---1.81)\\
  $\tau_4$&0. (0.---0.63)&0. (0.---0.64)&0. (0.---0.55)\\
  \hline
  RSS, $10^{-3}$&3.4&3.85&1.56\\
  \hline
\end{tabular}
\ec
\caption[]{Average turnover rates of CD8$^+$ T cells from four healthy humans as
  estimated by fitting the data from \citet{Mohri.jem01} using the
  Asymptote model, the Exponential model and the Gamma model. The best
  fits of the models resulted in different parameter estimates for all
  patients, with the exception of the fraction of turning over cells
  $\al$ in the Exponential model (which was fitted as one parameter
  for all patients).  As for CD4$^+$ T cells, in
  the model with gamma distributed turnover rates, the asymptote level
  $\al =1$ provided the best fit of the data.   
  The shown 95\% confidence intervals were obtained
  by bootstrapping the residuals with 1000 simulations.
}
\label{tab:mohri.controls.cd8}
\end{table}

\end{document}